\documentclass[conference]{IEEEtran}
\IEEEoverridecommandlockouts
\usepackage{cite}
\usepackage{amsmath,amssymb,amsfonts}
\usepackage{algorithmic}
\usepackage{graphicx}
\usepackage{epsfig}
\usepackage{epstopdf}
\usepackage{textcomp}
\usepackage{xcolor}
\usepackage{multirow}
\usepackage{amsmath}
\usepackage{subcaption}
\usepackage{fixltx2e}
\usepackage{float}
\usepackage{siunitx}
\usepackage{hyperref}
\hypersetup{colorlinks,urlcolor=blue}
\usepackage{soul}
\usepackage{colortbl}
\usepackage{authblk}
\usepackage[bottom,flushmargin]{footmisc}

\def\BibTeX{{\rm B\kern-.05em{\sc i\kern-.025em b}\kern-.08em
    T\kern-.1667em\lower.7ex\hbox{E}\kern-.125emX}}
\begin{document}

\title{Congestion Control in the Cellular-V2X Sidelink}

\author{Brian McCarthy}
\author{Aisling O'Driscoll}

\affil{{\textit{School of Computer Science and Information Technology}, \textit{University College Cork}, Cork, Ireland}\\\{b.mccarthy, a.odriscoll\}@cs.ucc.ie}

\maketitle
\begin{abstract}
This paper presents a detailed quantitative evaluation of standardised Decentralised Congestion Control (DCC) and packet dropping mechanisms for Cellular V2X (C-V2X). Based on the identified shortcomings, an Access layer DCC scheme, \textit{RRI adaptive}, is then proposed. \textit{RRI adaptive} accommodates the sidelink scheduling mechanism Sensing Based Semi-Persistent Scheduling (SB-SPS), eliminating incompatibilities between current standards and the scheduling mechanism, to avoid unnecessary and reoccurring collisions. Two variants are proposed; one is an evolution of the ETSI Reactive DCC mechanism and the other aligns with the 3GPP approach based on channel occupancy ratio (CR). Both approaches are compared with current ETSI and 3GPP standards and exhibit improved performance. An evaluation of existing DCC standards and \textit{RRI Adaptive} to meet the Quality of Service (QoS) requirements of vehicular cooperative awareness applications is also conducted.
\end{abstract} 


\section{Introduction}
\label{sec:intro}
Cooperative awareness between vehicles or between vehicles and infrastructure forms the basis for future envisaged vehicular communication services. There are currently two communication technologies i.e. the well studied and mature IEEE 802.11p (ITS-G5 in Europe) or the emerging cellular equivalent known as C-V2X. C-V2X defines two modes as part of the 3GPP (3rd Generation Partnership Project) Release 14 LTE-V standard \cite{3gpp-TR-36-885} and forming the basis for new radio (NR-V2X) in 3GPP Release 16 \cite{3gpp-rel16}. In Mode 3 (scheduled), the cellular base station allocates and manages the resources necessary for V2V PC5 communications. In Mode 4 (autonomous) each vehicle (VUE) selects its radio resources for V2V communications using the distributed scheduling algorithm SB-SPS. Mode 4 represents the baseline performance for C-V2X. 

Irrespective of whether wireless or cellular V2X technology is employed, both may be faced with handling a congested radio environment due to limited spectrum, widespread vehicular deployment and frequent packet exchange. As such, congestion control techniques are hugely important to manage channel load and radio interference. DCC mechanisms as defined by ETSI (European Telecommunications Standards Institute) have been well studied over the past decade for ITS-G5 \cite{limeric_paper, Rostami2016, Kenney2013} but have not been adequately investigated for C-V2X. The operation of the C-V2X MAC differs significantly from that of ITS-G5, particularly with respect to MAC layer scheduling. SB-SPS assumes packets arrive periodically and bases its resource reservation algorithm on this assumption. However, if packets arrive aperiodically in accordance with the ETSI CAM (Cooperative Awareness Message) generation rules or due some congestion control mechanism based on transmission rate control (TRC), this can lead to prohibitively high packet collisions. Furthermore, as SB-SPS maintains resources for a defined period known as a grant, these collisions may reoccur over the duration of the grant. Some initial studies \cite{cc-hari-missing-sci, Wendland2019} have investigated the application of packet dropping for SB-SPS but no study considers all ETSI and 3GPP DCC and packet dropping standards as in this paper. 

To address the identified shortcomings, this paper proposes Access layer DCC mechanisms that are compliant with the SB-SPS algorithm yet address the problem of excessive collisions due to the perception of unused yet reserved resources. This is achieved by adapting the resource reservation interval (RRI) within the SB-SPS grant in line with measured CBR (Channel Busy Ratio) as per the ETSI DCC mechanism or channel occupancy as per 3GPP. This simple approach drastically reduces the collisions by ensuring that resource usage is more predictable and does not result in reoccurring collisions within an SB-SPS grant. 

It is also important to recognise that while Access layer congestion control may improve radio conditions and in turn the measured packet delivery rates, it may not improve the performance achieved at the application layer. If a high number of application layer or cooperative awareness packets are delayed or dropped this may render the service unsafe or unusable. As such, it is of the utmost importance to consider the impact of Access layer DCC on the upper layer applications. To the author's knowledge, this paper is the first to do so for C-V2X and to investigate the impact of all DCC standards on these QoS requirements.

The rest of this paper is organised as follows: Section \ref{background} gives an overview of standardised ETSI and 3GPP DCC approaches as well as state of the art DCC research. Our proposed C-V2X DCC Access mechanisms that account for the C-V2X MAC layer are presented in Section \ref{sec:adaptive rri} and evaluated in Section \ref{sec:results}. Section \ref{sec:conclusion} provides concluding remarks.

\let\thefootnote\relax\footnote{This work has been submitted to the IEEE for possible publication. Copyright may be transferred without notice, after which this version may no longer be accessible.} 

\section{Congestion Control for Vehicular Communications}
\label{background}
\subsection{ETSI Decentralised Congestion Control (DCC)}
\label{subsec:congestion-ETSI}
ETSI defines a number of DCC mechanisms in \cite{ETSI2018}, originally developed for ITS-G5.
The de-facto mechanism is transmission rate control which works by increasing the delay between packet transmissions based on the CBR. The means by which this delay is determined is how ETSI distinguishes between it's TRC mechanisms, namely DCC Reactive and DCC Adaptive. 

DCC Reactive is the original approach specified by ETSI based on a state machine, where the state is associated with a particular CBR range. Depending on the CBR, a delay is introduced between consecutive packets to control the transmission rate. This is shown in Table \ref{dcc_access_reactive_table}. The maximum allowed transmission rate for a CBR range is enforced using T\textsubscript{off}, the time period before a new consecutive packet can be transmitted. DCC Adaptive, is a rate control mechanism based on the LIMERIC algorithm \cite{limeric_paper}. It uses an algorithmic approach to calculating the delay rather than a predefined lookup table. It is designed to adjust packet rate transmission to converge on a target CBR, with a default of 68\%. DCC Adaptive better considers factors such as stability and fairness to ensure no node is starved. Amador et. al. provide a detailed analysis of ETSI DCC Adaptive in \cite{Amador2020}.

\begin{table}[htbp]
\caption{ETSI Access Reactive DCC as specified in \cite{ETSI2018}.}
\label{dcc_access_reactive_table}
\begin{center}
\begin{tabular}{l l l l}
\hline\hline 
\textbf{CBR} & \textbf{State} & \textbf{Packet  Rate} & \textbf{T\textsubscript{off}} \\ [0.5ex] 
\hline
CBR $<$ 0.3 & Relaxed & 10Hz & 100ms\\
\hline 
0.3 $\leq$ CBR $\leq$ 0.40 & Active 1 & 5Hz & 200ms\\
\hline
0.40 $\leq$ CBR $\leq$ 0.50 & Active 2 & 2.5Hz & 400ms\\
\hline
0.50 $\leq$ CBR $\leq$ 0.60 & Active 3 & 2Hz & 500ms\\
\hline
CBR $>$ 0.60 & Restrictive & 1Hz & 1000ms\\
\hline\hline
\end{tabular}
\end{center}
\end{table} 

\subsection{ETSI C-V2X Congestion Control}
\label{subsec:congestion-CV2X}
In recent years, ETSI have set out some details relating to congestion control for C-V2X \cite{ETSI-cellular}. Specifically, they describe how CBR and CR are to be measured. CBR provides an estimation of the overall channel congestion by measuring the ratio of subchannels over the last 100 subframes where the sidelink RSSI (S-RSSI) exceeds a predefined threshold. The CR measures the number of subchannels used by each vehicle over a historical time period as well as the number that will be used based on the current configured grant. In Table \ref{etsi_cbr_cr_table} ETSI specifies the maximum CR limit for each vehicle based on the measured CBR. If the measured CR exceeds the limit for the current CBR range, the vehicle must reduce it's CR using a particular congestion control mechanism. ETSI highlights that this can include packet dropping, MCS adaptation and power control but do not specify precise implementations. 

\begin{table}[htbp]
\caption{ETSI C-V2X Congestion Control - Maximum CR limit per CBR range and packet priority \cite{ETSI-cellular}.}
\label{etsi_cbr_cr_table}
\begin{center}
\begin{tabular}{l l l l}
\hline\hline 
\textbf{CBR} & \textbf{Priority 1-2} & \textbf{Priority 3-5} & \textbf{Priority 6-8} \\ [0.5ex] 
\hline
0 $\leq$ CBR $\leq$ 0.3 & no limit & no limit & no limit\\
\hline 
0.3 $<$ CBR $\leq$ 0.65 & no limit & 0.03 & 0.02\\
\hline
0.65 $<$ CBR $\leq$ 0.8 & 0.02 & 0.006 & 0.004\\
\hline
0.8 $<$ CBR $\leq$ 1 & 0.002 & 0.003 & 0.002\\
\hline\hline
\end{tabular}
\label{tab:road}
\end{center}
\end{table} 

\subsection{3GPP C-V2X Congestion Control}
\label{subsec:3gppPacketDropping}

The 3GPP C-V2X standard also does not provide a specific congestion control mechanisms but, similarly to ETSI, defines the CBR and CR measurements under which it should be invoked. It also specifies a table shown in Table \ref{qualcomm_cbr_cr_table}, as part of a 3GPP working group, which is less reactive than the ETSI equivalent at lower CBR levels.For the remainder of this paper we assume packet dropping is the congestion control mechanism invoked by this table. 

\begin{table}[htbp]
\caption{3GPP C-V2X Congestion Control - Maximum CR limit per CBR range and packet priority \cite{Qualcomm2016}.}
\label{qualcomm_cbr_cr_table}
\begin{center}
\begin{tabular}{l l l l}
\hline\hline 
\textbf{CBR} & \textbf{CR limit} \\ [0.5ex] 
\hline
CBR $\leq$ 0.65 & no limit\\
\hline 
0.65  $<$ CBR $\leq$ 0.675 & 1.6e-3\\
\hline
0.675 $<$ CBR $\leq$ 0.7 & 1.5e-3\\
\hline
0.7   $<$ CBR $\leq$ 0.725 & 1.4e-3\\
\hline
0.725 $<$ CBR $\leq$ 0.75 & 1.3e-3\\
\hline
0.75  $<$ CBR $\leq$ 0.775 & 1.2e-3\\
\hline
0.8   $<$ CBR $\leq$ 0.825 & 1.1e-3\\
\hline
0.825 $<$ CBR $\leq$ 0.85 & 1.1e-3\\
\hline
0.85  $<$ CBR $\leq$ 0.875 & 1.0e-3\\
\hline
0.875 $<$ CBR & 0.8e-3\\
\hline\hline
\end{tabular}
\end{center}
\end{table}

\subsection{Literature Review}
\label{subsec:stateofart}

One of the initial works to evaluate the impact of packet dropping for C-V2X is by Mansouri et. al \cite{Mansouri2019}. It evaluates the performance of packet dropping based on the 3GPP C-V2X CR limits as per Table \ref{qualcomm_cbr_cr_table}. This paper highlights some of the issues that we discuss in Section \ref{subsec:adaptiveRRI-results} where VUEs mistakenly choose the same channel resources. The most similar work to the approach proposed in this paper is a reservation splitting technique by Wendland et. al \cite{Wendland2019}. The authors split a single SB-SPS grant into multiple sub-grants of lower frequency (e.g. 10Hz $\rightarrow$ 2 x 5Hz) and  when the network is congested, individual sub-grants can be disabled without interrupting the SB-SPS grant mechanism. This approach is analogous to the proposed \textit{RRI Adaptive} mechanism in that turning off a grant is similar to changing the RRI i.e. a single 5hz grant is the same as a grant with an increased RRI of 200ms. They compare their scheme against ETSI C-V2X DCC as per Table \ref{etsi_cbr_cr_table}. While the reservation splitting approach reduces recurring collisions within a single grant, it does not support dynamically re-enabling grants as congestion changes. In contrast the proposed \textit{RRI Adaptive} approaches allow this to occur and work within the existing SB-SPS mechanism without requiring any changes. Sepulcre et. al. \cite{Sepulcre2020} recently investigated the efficacy of packet dropping to meet application QoS requirements as a congestion control mechanism for ITS-G5. The authors show that application PDR is impacted by packet dropping and performs worse than if no congestion control was applied despite improvements in radio performance. However, it is our premise that packet dropping can form part of a congestion control solution for C-V2X as seen in Fig \ref{fig:IPG-60pc} and Table \ref{tab:Awareness} where we observe similar IPG and increased neighbour awareness. This also highlights a need for further investigation into whether the transmission frequency of CAMs and other cooperative awareness services can be increased such that they are only transmitted when meaningful and without impacting the awareness of vehicles. A similar premise was recently discussed by Bazzi et. al \cite{BazziDoICare2018}.
 
Other approaches look at combining mechanisms such as rate control, power control and MCS adaptation. The most prevalent of these is the SAE DCC mechanism \cite{SAE-standard} that uses power and rate control. The rate control algorithm is derived from the Limeric algorithm \cite{limeric_paper} and power control is based on The Stateful Utilization-
based Power Adaptation (SUPRA) which is designed to control communication range \cite{supra}. Research in \cite{toghi-dcc-first,Toghi-dcc-spatio,YoonPowerRate,ChoiQOS} investigate the performance of the SAE standard for congestion control. Generally, these authors have shown performance increases over standard rate control while showing minor improvements from power control, with the need for further study before determining their effectiveness for C-V2X.

Notably, all the mechanisms described thus far are Access layer approaches. Such mechanisms have limitations when simultaneously considering multiple application layer services with different QoS requirements. As such, ETSI has begun defining the Facilities layer DCC which coordinates between Access layer congestion control mechanisms and higher layer services. It can also operate as a standalone mechanism without lower layers. The goal is to ensure that nodes can predict available resources and more intelligently distribute channel resources across the services that they are tasked with fulfilling. ETSI has not yet standardised Facilities layer DCC but some works \cite{Khan2018,Khan2020,Delooz2020,Amador2020} have investigated possible implications and implementations.

\section{Proposed Adaptive RRI Approaches}
\label{sec:adaptive rri}

In Section \ref{sec:results} it is shown that traditional TRC mechanisms like DCC or packet dropping are incompatible with the SB-SPS mechanism when a scheduled transmission is missed, leading to unnecessary collisions. To overcome this, two \textit{adaptive RRI} mechanisms are proposed. These mechanisms adjust the time between transmissions based on the current CBR measurement, while considering the RRI of the SB-SPS grant. Two variants are proposed;

\begin{itemize}
    \item \textbf{RRI\textsubscript{lookup}}: This is based on a TRC mechanism similar to DCC Reactive, which uses a lookup table to determine the packet delay i.e. T\textsubscript{off}, based on the measured CBR. However in this case the delay offset is linked to a multiple of the default RRI and an SCI is broadcast indicating the new RRI. This controls the rate of packet transmission without missing a scheduled resource reservation slot. 
    \item \textbf{RRI\textsubscript{CR\_limit}}: This approach is based on packet dropping to reduce an individual VUE's CR-limit. This is similar to the ETSI and 3GPP C-V2X approaches in that it is also based on a CBR CR limit table. Given a particular CBR, representing the overall measured congestion of the channel, if it is determined that the VUE's CR exceeds the CR limit for that CBR, the RRI will be increased such that the CR is brought below the limit. An SCI is then broadcast indicating the new RRI. When the CBR returns to a lower range and CR can be increased, a new RRI will is chosen that maintains the new limit.
\end{itemize}

\begin{figure}[htbp]
  \centering
  \includegraphics[width=.82\linewidth]{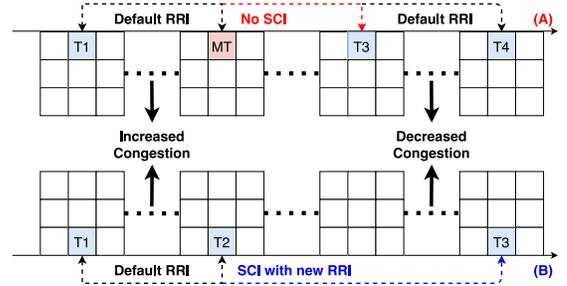}
\caption{RRI Adaptive approaches versus previous
\label{fig:rri-adapt-diagram}approaches.}
\end{figure}

As both of these approaches operate at the MAC layer, they represents Access layer congestion control mechanisms.  We also employ the DCC averaging mechanism where RRI transitions only occur after 1 second of CBR exceeding a threshold and RRIs are decreased after 5 seconds of lower measured CBR. This results in a more stable level of CBR for all nodes. Fig. \ref{fig:rri-adapt-diagram} illustrates the concept underpinning both \textit{adaptive RRI} approaches (shown as per timeline \textit{B}) with the default SB-SPS operation shown as per timeline \textit{A}. In \textit{A}, after transmission \textit{T1}, congestion occurs. This results in a delay by a TRC congestion control mechanism or a packet drop. The consequence is that a packet is not transmitted in the next scheduled resource reservation slot i.e. a missed transmission (\textit{MT}). As a result, an SCI is not transmitted so neighbouring VUEs believe the resource(s) to be free. However as the SB-SPS grant is maintained, and may be utilised in the future e.g. transmission \textit{T3}, a collision on that resource can occur. In contrast, the proposed \textit{adaptive RRI} mechanisms shown in \textit{B}, explicitly considers the RRI in the grant when adding a delay in the case of \textit{RRIlookup} or packet dropping in the case of \textit{CR limit}. When transmission \textit{T1} occurs it is determined the new RRI and an SCI is broadcast with the new RRI to all neighbouring VUEs. This ensures that neighbouring VUEs are informed of when the vehicle next intends to transmit and can utilise the free resources in the interim. 

\section{Results}
\label{sec:results}

The road environment, vehicular density and other key simulation parameters used in this evaluation are summarised in Table \ref{tab:cv2x-setup}. This is designed to allow for comparison with a previous study \cite{cc-hari-missing-sci}. Application packets are generated periodically (100ms) but only the most recent packet is transmitted i.e. older intermediate packets are dropped if a newer packet is generated.

\begin{table}[htbp]
\caption{Simulation Parameters.}
\begin{center}
\begin{tabular}{l l}
\hline\hline
\textbf{Parameter} & \textbf{Value}\\
\hline\hline
\multicolumn{2}{c}{\textbf{Vehicular scenario}}\\
\hline
Vehicular density & 0.46 veh/m\\
Road length & 600 m \\
Number of lanes & 3 in each direction (6 in total)\\
Vehicle Speed & 50km/h\\
SUMO step-length & 1ms\\
\hline
\multicolumn{2}{c}{\textbf{Channel settings}}\\
\hline
Carrier frequency & 5.9 GHz\\
Channel bandwidth, \/No. subchannels & 10 MHz, 3\\
Subchannel size & 16 Resource Blocks\\
\hline
\multicolumn{2}{c}{\textbf{Application layer}}\\
\hline
Packet size & 190 Bytes\\
Transmission frequency ($F_{Tx}$) & 10 Hz\\
\hline
\multicolumn{2}{c}{\textbf{MAC \& PHY layer}}\\
\hline
Resource keep probability & 0\\
RSRP threshold & -126 dBm \\
RSSI threshold & -90 dB \\
Propagation model & Winner+ B1\\
MCS & 6 (QPSK 0.5)\\
Transmission power ($P_{Tx}$) & 23 dBm\\
Noise figure & 9 dB\\
Shadowing variance LOS & 3 dB\\
\hline\hline
\end{tabular}
\label{tab:cv2x-setup}
\end{center}
\end{table}

\subsection{ETSI Access layer Congestion Control}
\label{sec:currentSolutions}

\begin{figure}[htbp]
  \centering
  \includegraphics[width=.82\linewidth]{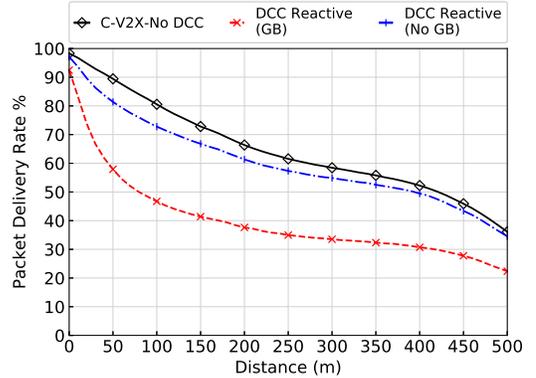}
\caption{ETSI Access layer DCC (Reactive).}
\label{fig:dcc-gb-issue-pdr}
\end{figure}

We investigate the efficacy of the standardised ETSI DCC mechanism, designed for ITS-G5, to the 3GPP C-V2X standard, and compare to it's performance without congestion control (labelled \textit{C-V2X-No DCC}). Fig. \ref{fig:dcc-gb-issue-pdr} shows the PDR, considering two C-V2X SB-SPS configurations. The first assumes the default configuration where grant breaking is enabled (labelled \textit{DCC Reactive (GB)}) and with the \textit{sl-reselectAfter} parameter set to 1. This means that the grant is broken if a single reserved resource is not used i.e. other vehicles will perceive future grant resources to be free. The second configuration assumes grant breaking is disabled (labelled \textit{DCC Reactive (no GB)}).

We can observe that DCC has a large detrimental impact on the PDR performance of C-V2X when grant breaking is enabled. This is for the same reasons that cause the sharp decline of SB-SPS when scheduling aperiodic application traffic, as discussed in detail in \cite{mccarthy2021opencv2x}.
As described in Section \ref{subsec:congestion-ETSI}, DCC introduces a delay in the packet inter-arrival time (mean of 300ms). If the packet inter-arrival time increases beyond a maximum of 198ms, assuming a resource reservation interval (RRI) of 100ms (\textit{2n-2} where \textit{n}=RRI)\cite{mccarthy2021opencv2x}, this breaks the grant. Grant breaking leads to a rise in collisions due to VUEs contending for an increasingly small CSR pool. This is compounded by constant rescheduling, increasing the risk of VUEs selecting the same resource, thereby decreasing the overall PDR. To alleviate this, grant breaking can be disabled which improves performance due to reduced resource rescheduling, as shown in Fig. \ref{fig:dcc-gb-issue-pdr}. However \textit{DCC Reactive (No GB)} also incurs a degradation in PDR when compared with \textit{C-V2X-No DCC}. This is because disabling grant breaking introduces an additional source of error, as discussed by Harri et al \cite{cc-hari-missing-sci}. If reserved resources go unused but the grant is maintained, an SCI is not sent. Thus, neighbouring VUEs may mistakenly believe the resources to be free, leading to unnecessary collisions. For the remainder of the paper we will consider \textit{DCC  Reactive (No GB)} as a baseline, herein referred to as \textit{DCC  Reactive}. 

We next compare \textit{DCC Reactive} against the other defined approach in the ETSI standard, namely \textit{DCC Adaptive}. We consider default \textit{DCC Adaptive} behaviour where the CBR converges to a target of 68\%. We also evaluate with a CBR target of 20\%, comparable to the mean CBR that was measured for \textit{DCC Reactive}, to allow for direct comparison. 

\begin{figure}[htbp]
\begin{subfigure}{.48\textwidth}
  \centering
  \includegraphics[width=.82\linewidth]{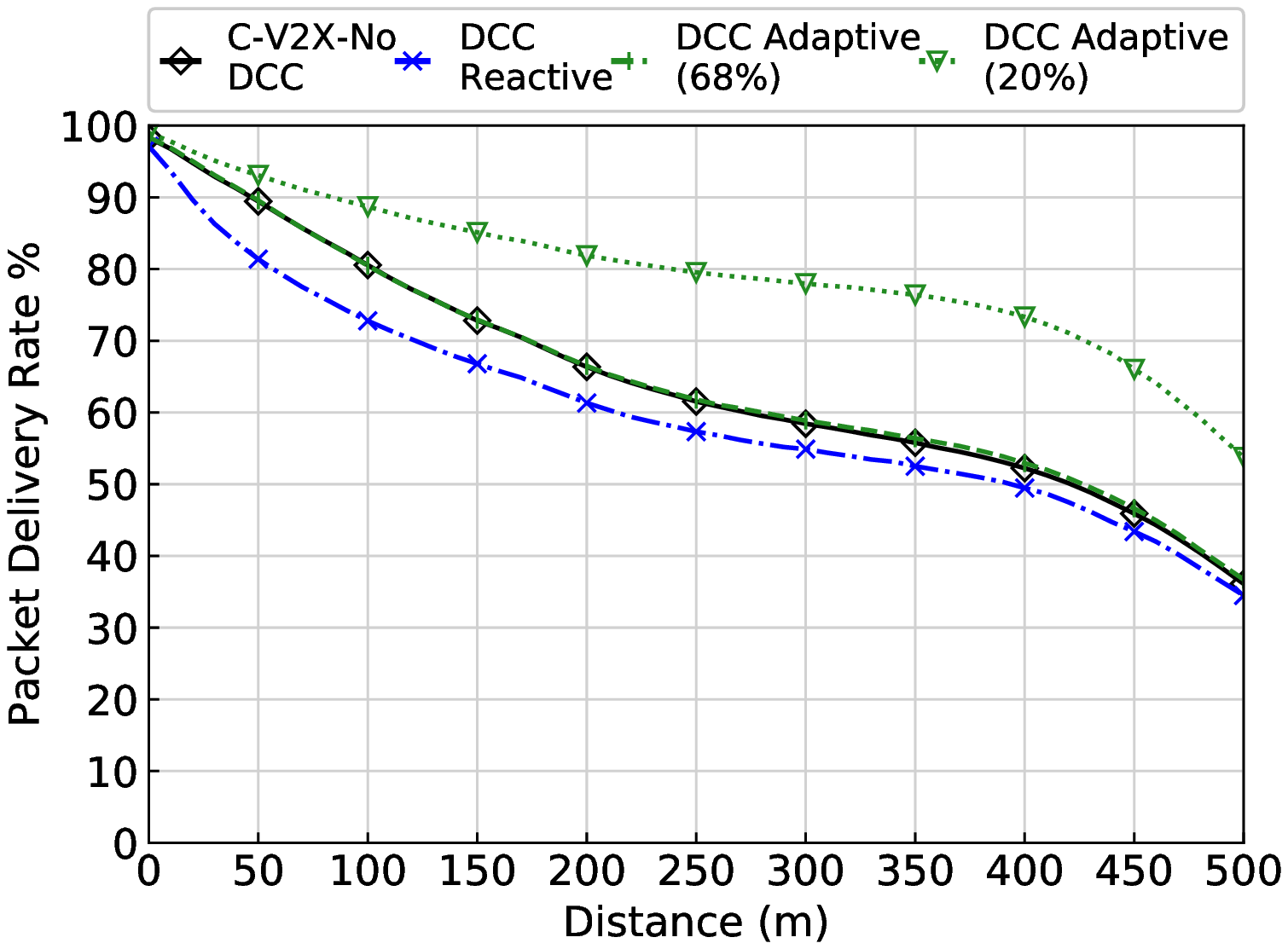}
  \caption{ETSI DCC PDR.}
  \label{fig:dcc-variants-pdr}
\end{subfigure}
\begin{subfigure}{.48\textwidth}
  \centering
  \includegraphics[width=.82\linewidth]{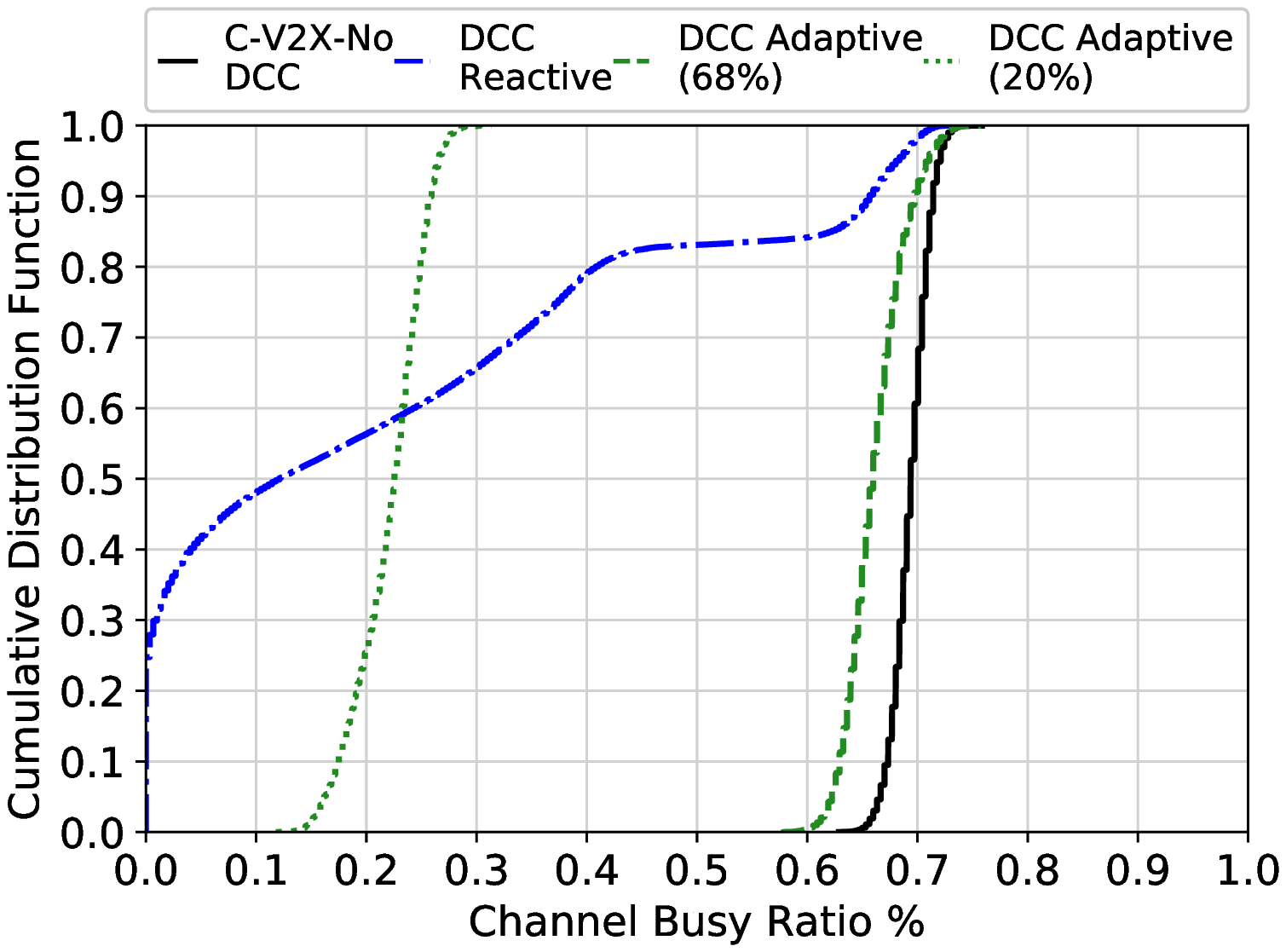}
  \caption{ETSI DCC CBR.}
  \label{fig:dcc-variants-cbr}
\end{subfigure}
\caption{Impact of DCC variants on PDR \& CBR (No grant breaking).}
\label{fig:dcc-variants}
\end{figure}

The impact on PDR is shown in Fig. \ref{fig:dcc-variants-pdr}. Notably \textit{DCC Adaptive} (68\%) incurs similar performance to no congestion control due to comparable CBR, as shown in Fig. \ref{fig:dcc-variants-cbr}. 
Of greater interest, is the comparison between \textit{DCC Adaptive} and \textit{DCC Reactive} for a comparable CBR of 20\%. DCC reactive performs poorly for the reasons already discussed and due to drawbacks associated with stability \cite{Rostami2016}. It is notable that when CBR is reduced to a target of 20\%, the performance of \textit{DCC Adaptive} greatly improves PDR by up to 21\%. Thus the performance of \textit{DCC Adaptive} is further explored in comparison with alternative mechanisms in Section \ref{subsec:adaptiveRRI-results}.

\subsection{ETSI \& 3GPP Packet Dropping}
\label{subsec:results-packetDropping}

The limitations of the ETSI DCC mechanisms when applied to C-V2X have been shown, caused by incompatibility with the default SB-SPS algorithm. ETSI and 3GPP congestion control recommendations are next evaluated with an emphasis on packet dropping. As described in \ref{subsec:congestion-CV2X} and \ref{subsec:3gppPacketDropping}, this method of congestion control simply drops packets before transmission to maintain a specific CR limit according to a lookup table. Three CR limit lookup tables are considered:

\begin{itemize}
    \item The recently defined ETSI CR limit table as shown in Table \ref{etsi_cbr_cr_table}. A traffic priority of 6-8 is assumed as other priorities had negligible impact. This is labelled \textit{Packet Dropping (ETSI)}.
    \item As part of a 3GPP working group, Qualcomm have also proposed a CR limit table \cite{Qualcomm2016}. This is labelled \textit{Packet Dropping (3GPP)}.
    \item An adapted 3GPP CR limit table, with lower CBR thresholds i.e. designed to more aggressively manage congestion and to produce a mean CBR closer to 20\% in line with the \textit{DCC Reactive} scheme. This is labelled \textit{Packet Dropping (Aggressive)}.
\end{itemize}

\begin{figure}[htbp]
\begin{subfigure}{.48\textwidth}
  \centering
  \includegraphics[width=.82\linewidth]{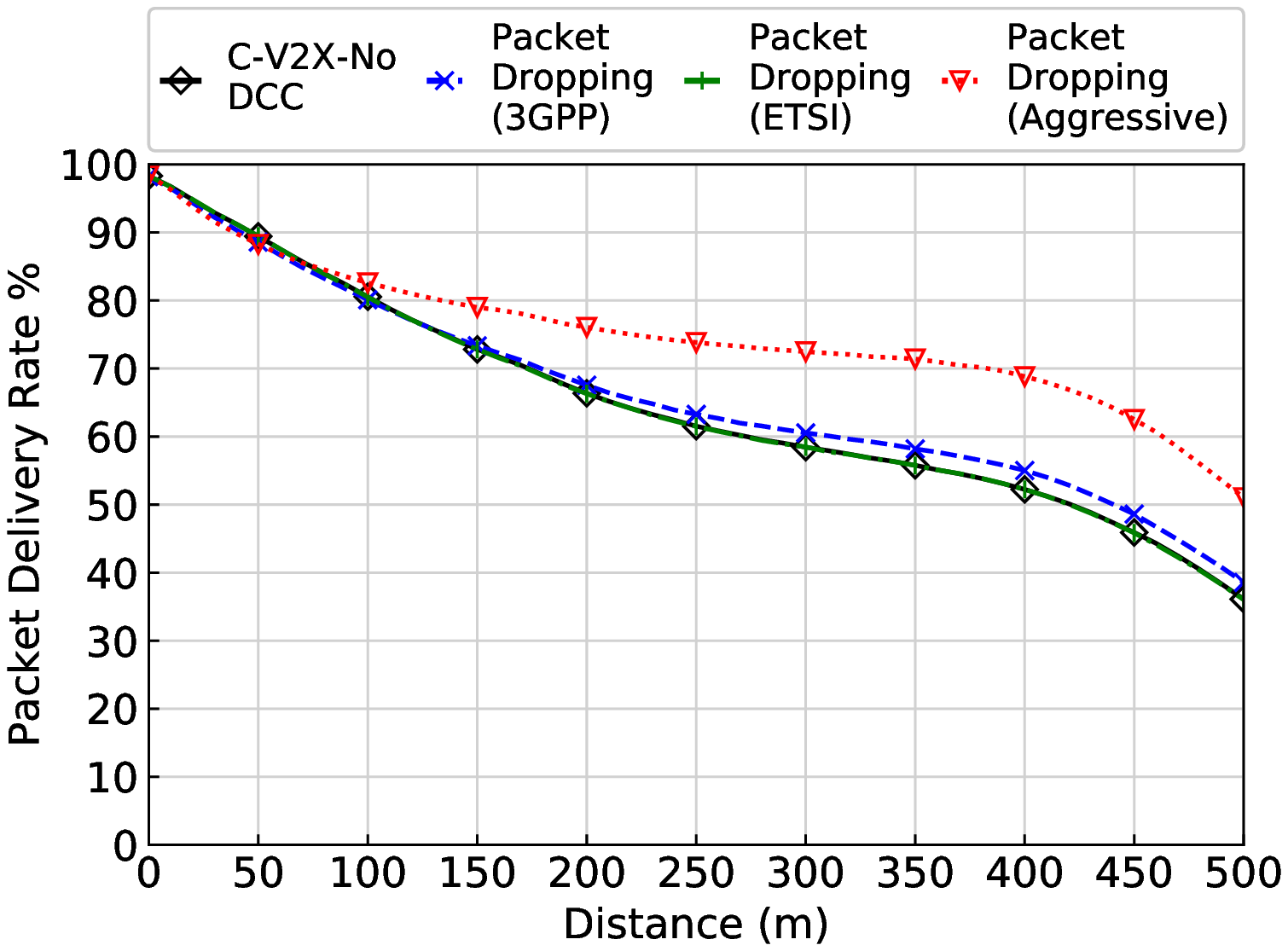}
  \caption{Packet Dropping PDR.}
  \label{fig:pdrop-pdr}
\end{subfigure}
\begin{subfigure}{.48\textwidth}
  \centering
  \includegraphics[width=.82\linewidth]{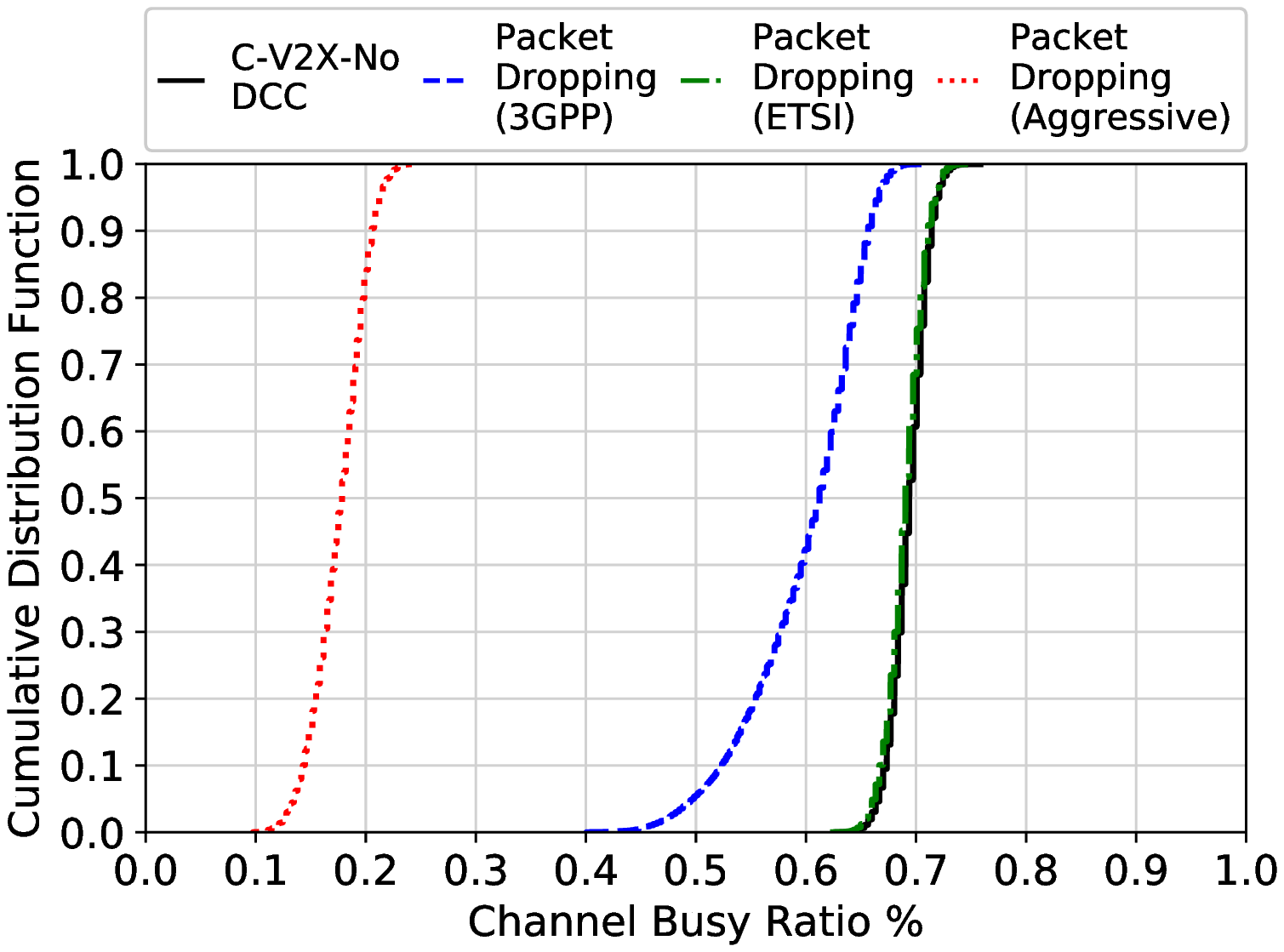}
  \caption{Packet Dropping CBR.}
  \label{fig:pdrop-cbr}
\end{subfigure}
\caption{ETSI \& 3GPP Packet Dropping.}
\label{fig:pdrop}
\end{figure}

The PDR performance can be seen in Fig. \ref{fig:pdrop-pdr}. \textit{Packet Dropping (ETSI)} performs identically to no congestion control both in terms of PDR and CBR. \textit{Packet Dropping (3GPP)} performs marginally better due to a reduced CBR of approx. 10\%, as shown in Fig. \ref{fig:pdrop-cbr}, although this only translates into a PDR improvement of 3\% at distances exceeding 150m as seen in Fig. \ref{fig:pdrop-pdr}. Adopting a more aggressive approach to reducing CBR in the case of \textit{Packet Dropping (Aggressive)}, improves PDR by up to 17\% at distances beyond 100m. However, the improvement in PDR does not correlate with the significantly lower CBR.

\subsection{Adaptive RRI  - Access layer Performance}
\label{subsec:adaptiveRRI-results}

Based on the drawbacks of existing standardised approaches, we next evaluate the performance of the proposed \textit{adaptive RRI} congestion control schemes as described in Section \ref{sec:adaptive rri}. These are compared against \textit{C-V2X-No DCC}, \textit{DCC Reactive}, \textit{DCC Adaptive} (CBR target of 20\%) and \textit{Packet Dropping (Aggressive)}. In all cases we consider grant breaking to be disabled. 

\begin{table*}[t]
\caption{Absolute number of colliding grants.}
\begin{center}
\begin{tabular}{c l c c c c c}
\hline\hline
\textbf{CBR} &
\textbf{Congestion Control Mechanism} &
\textbf{Mean Colliding Grants ($\gamma$)} & 
\textit{\(\gamma\)\textsubscript{MT}} (\%) & 
\textit{\(\gamma\)\textsubscript{NF}} (\%) &
\textit{\(\gamma\)\textsubscript{TSim}} (\%)\\
\hline
\multirow{6}{*}{20\%} & \textit{C-V2X-No DCC} & 5076 & - & 3645 & 1431\\
\multirow{6}{*}{} & \textit{DCC Adaptive} & 2286 & 678 & 0 & 1608\\
\multirow{6}{*}{} & \textit{DCC Reactive} & 4885 & 338 & 399 & 4148\\
\multirow{6}{*}{} & \textit{Packet Dropping (Aggressive)} & 1565 & 889 & 0 & 676\\
\multirow{6}{*}{} & \textit{RRI\textsubscript{Lookup}} & 427 & - & 258 & 169\\
\multirow{6}{*}{} & \textit{RRI\textsubscript{CRlimit}} & 62 & - & 0 & 62\\\hline
\multirow{2}{*}{60\%} & \textit{DCC Adaptive} & 4954 & 842 & 2626 & 1486\\
\multirow{2}{*}{} & \textit{RRI\textsubscript{CRlimit}} & 768 & - & 558 & 210 \\\hline\hline
\end{tabular}%
\label{tab:colliding_grants}
\end{center}
\end{table*}

\begin{figure}[htbp]
\begin{subfigure}{.48\textwidth}
\centerline{
\includegraphics[width=.82\linewidth]{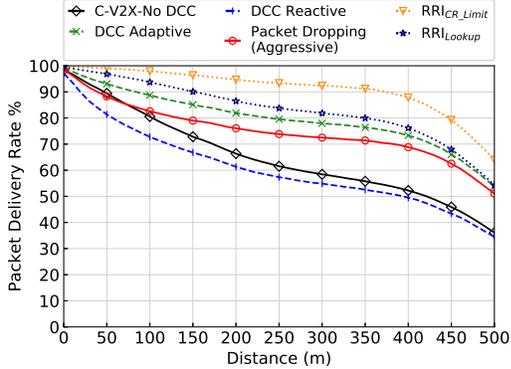}}
\caption{Packet Delivery Rate.}
\label{fig:RRI-PDR-20pc}
\end{subfigure}
\begin{subfigure}{.48\textwidth}
\centerline{
\includegraphics[width=.82\linewidth]{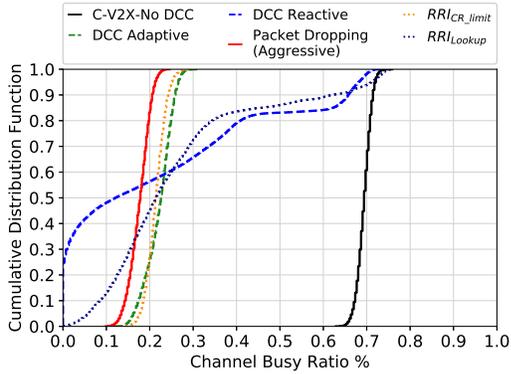}}
\caption{Channel Busy Ratio.}
\label{fig:RRI-CBR-20pc}
\end{subfigure}
\caption{Comparative congestion control at 20\% CBR.}
\label{fig:RRI-20pc}
\end{figure}

Fig. \ref{fig:RRI-PDR-20pc} shows the PDR performance of \textit{RRI\textsubscript{lookup}} outperforms \textit{DCC Adaptive} by up to 5\% PDR but Fig. \ref{fig:RRI-CBR-20pc} demonstrates less stability with respect to CBR. This is because it inherits the instability of a table lookup mechanism such as \textit{DCC Reactive}. It exhibits considerably less mean colliding grants when compared to standardised DCC approaches as shown in Table \ref{tab:colliding_grants}. More notably, \textit{RRI\textsubscript{CR\_limit}} shows significantly improved PDR over all schemes (a mean of 11\% when compared to \textit{DCC Adaptive}) even those with comparable CBR. It also exhibits the lowest number of mean colliding grants.

The mean colliding grants is a significant metric as it represents the resources utilised by 2 or more VUEs simultaneously, due to poor CSR selection in SB-SPS. Colliding grants can occur for three reasons:

\begin{enumerate}
    \item A missed transmission causing a VUE to not send an SCI. Neighbouring VUEs will mistakenly believe a resource to be free in future reserved slots, when they may be utilised. This is denoted as \textit{$\gamma$\textsubscript{MT}} in Table \ref{tab:colliding_grants}.
    \item In congested radio conditions, when no resources are determined to be free, VUEs will select resources with low RSRP and RSSI measurements. This is denoted as \textit{$\gamma$\textsubscript{NF}} in Table \ref{tab:colliding_grants}.
    \item When neighbouring VUEs reserve the same resources within a single RRI as a result of highly similar selection windows. As neighbouring VUEs may have similar historical RSRP/RSSI measurements they are likely to have similar CSRs which can result in selecting the same resource(s). This is denoted as \textit{$\gamma$\textsubscript{TSim}} in Table \ref{tab:colliding_grants}.
\end{enumerate}

DCC and packet dropping mechanisms are very susceptible to colliding grants due to \textit{$\gamma$\textsubscript{MT}}, because of the incompatibility of their schemes with the SB-SPS RRI mechanism. This is evident in Table. \ref{tab:colliding_grants}. Also, as the reserved resources are maintained for 5-15 sub-frames, as determined by the Resource Reselection Counter (RRC), recurring colliding grants represent multiple avoidable half duplex errors in the channel as well as increasing interference. In contrast, the \textit{RRI Adaptive} approaches account for this and hence incur significantly less colliding grants by eliminating \textit{$\gamma$\textsubscript{MT}} and reducing \textit{$\gamma$\textsubscript{NF}} and \textit{$\gamma$\textsubscript{TSim}}. The \textit{RRI\textsubscript{lookup}} CBR variance results in a higher number of colliding grants than the more stable \textit{RRI\textsubscript{CR\_limit}}.

\begin{figure}[htbp]
\begin{subfigure}{.48\textwidth}
\centerline{
\includegraphics[width=.82\linewidth]{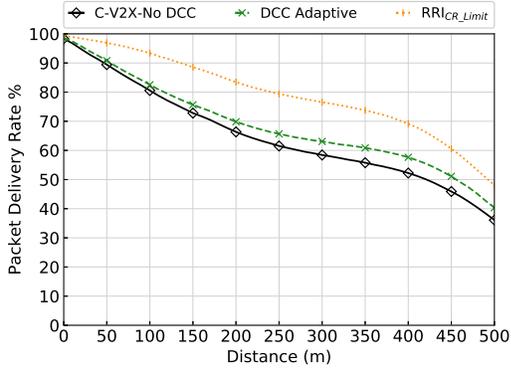}}
\caption{Packet Delivery Rate.}
\label{fig:RRI-PDR-60pc}
\end{subfigure}
\begin{subfigure}{.48\textwidth}
\centerline{
\includegraphics[width=.82\linewidth]{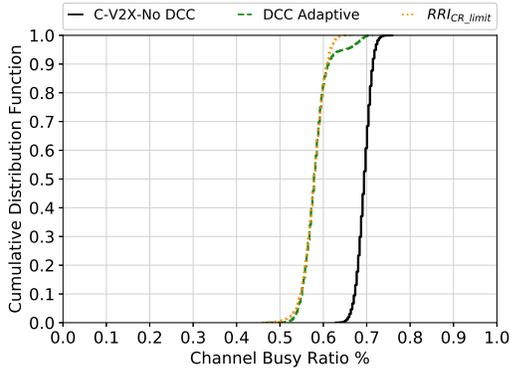}}
\caption{Channel Busy Ratio.}
\end{subfigure}
\label{fig:RRI-CBR-60pc}
\caption{Comparative congestion control at 60\% CBR.}
\label{fig:RRI-60pc}
\end{figure}

While \textit{RRI\textsubscript{CR\_limit}} demonstrates significantly higher PDR (Fig. \ref{fig:RRI-PDR-20pc}), it incurs comparable application performance to \textit{DCC Adaptive}, discussed further in Section \ref{subsec:appqos}. This is attributable to aggressive congestion control with CBR of approx. 20\% (for comparison with schemes such as \textit{DCC Reactive}). Given that this can result in under utilisation of the channel, a CBR of approximately 60\% is considered in Fig. \ref{fig:RRI-60pc}. This aligns with previous ITS-G5 studies that have investigated maximum throughput at 60\% \cite{Kenney2013}. \textit{RRI\textsubscript{CR\_limit}} is compared against \textit{C-V2X-No DCC} and \textit{DCC Adaptive}, which is the closest in performance. \textit{RRI\textsubscript{CR\_limit}} exhibits a significant PDR improvement of up to 18\% and 14\% respectively. 

\subsection{Adaptive RRI  - Application Layer Performance}
\label{subsec:appqos}

It is important to consider the performance of the proposed congestion control mechanisms for the applications that they service, particularly as vehicular communications often has geotemporal relevance. Specifically, we consider the impact on the Inter-Packet Gap (IPG), which represents the average elapsed time between receptions from neighbouring VUEs. We also consider mean awareness which is the percentage of vehicles a VUE is aware of in a given communication range. The maximum lifetime of a CAM before it is discarded by a VUE is 1 second.

\begin{figure}[htbp]
\begin{subfigure}{.48\textwidth}
\centerline{
\includegraphics[width=.82\linewidth]{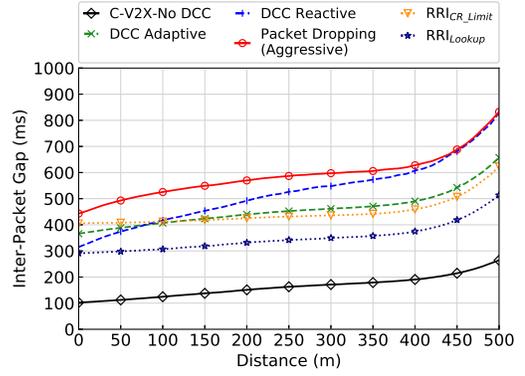}}
\caption{IPG at 20\% CBR.}
\label{fig:IPG-20pc}
\end{subfigure}
\begin{subfigure}{.48\textwidth}
\centerline{
\includegraphics[width=.82\linewidth]{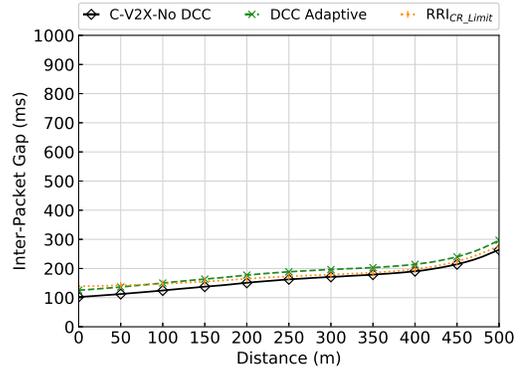}}
\caption{IPG at 60\% CBR.}
\label{fig:IPG-60pc}
\end{subfigure}
\caption{Inter-Packet Gap (IPG) performance of congestion control mechanisms.}
\label{fig:IPG}
\end{figure}

In Fig. \ref{fig:IPG-20pc}, \textit{RRI\textsubscript{lookup}} demonstrates the lowest IPG while exhibiting better PDR (Fig. \ref{fig:RRI-PDR-20pc}), except for \textit{RRI\textsubscript{CR\_limit}}. It demonstrates higher mean neighbour awareness than current standardised approaches as shown in Table \ref{tab:Awareness}, but lower than \textit{DCC Adaptive} and \textit{RRI\textsubscript{CR\_limit}}. The IPG demonstrated by \textit{RRI\textsubscript{CR\_limit}} is similar to that of \textit{DCC Adaptive} but exhibits much higher PDR. As shown in Table \ref{tab:Awareness}, it also demonstrates high neighbour awareness. Application layer performance for a CBR of 60\% is also considered. In Table \ref{tab:Awareness}, \textit{RRI\textsubscript{CR\_limit}} with aggressive does little to impact neighbour awareness when compared to 60\%. It does however impact the IPG. \textit{RRI\textsubscript{CR\_limit}} incurs higher IPG at 20\% CBR as the level of congestion control is very aggressive. Fig. \ref{fig:IPG-60pc} shows the IPG for 60\% CBR, with a mean of 165ms for \textit{C-V2X-No DCC}, 180ms for \textit{RRI\textsubscript{CR\_limit}} and 190ms for \textit{DCC Adaptive} respectively. 

\begin{table}[htbp]
\caption{Neighbour VUE Awareness (200m-300m)}
\label{tab:Awareness}
\begin{center}
\resizebox{\columnwidth}{!}{%
\begin{tabular}{c l c c}
\hline\hline 
 \textbf{CBR} &\textbf{Congestion Control Mechanism} & \textbf{Awareness \%}  & \textbf{Std. Dev. \%}\\ [0.5ex] 
\hline
\multirow{6}{*}{20\%} & \textit{C-V2X-No DCC} & 92.22 & 4.43\\
\multirow{6}{*}{} & \textit{DCC Adaptive} & 94.85 & 3.46 \\
\multirow{6}{*}{} & \textit{DCC Reactive} & 74.55 & 9.76 \\
\multirow{6}{*}{} & \textit{Packet Dropping (Aggressive)} & 80.41 & 5.58 \\
\multirow{6}{*}{} & \textit{RRI\textsubscript{Lookup}} & 90.91 & 4.38\\
\multirow{6}{*}{} & \textit{RRI\textsubscript{CR\_limit}} & 97.94 & 2.13 \\
\hline
\multirow{2}{*}{60\%} & \textit{DCC Adaptive} & 97.80 & 2.72\\
\multirow{2}{*}{} & \textit{RRI\textsubscript{CR\_limit}} & 97.75 & 2.24 \\
\hline\hline
\end{tabular}
}
\end{center}
\end{table}

\section{Conclusion}
\label{sec:conclusion}

In this paper a new DCC approach, \textit{RRI adaptive}, is proposed which is fully compatible with the SB-SPS mechanism. We have shown that \textit{RRI adaptive} enables much higher performance in terms of PDR while maintaining lower IPG and higher mean neighbour awareness than existing standardised approaches. Future work will seek to modify \textit{RRI adaptive} to avoid a table based approach which is prone to CBR instability. Open research questions also exist relating to Facilities layer congestion control that considers multiple simultaneous V2X services with diverse QoS requirements.

\section*{Acknowledgements}
This publication has emanated from research conducted with the financial support of Science Foundation Ireland (SFI) under Grant No: 17/RC-PhD/3479.

\bibliographystyle{IEEEtran}
\bibliography{references}

\begin{thebibliography}{10}
\providecommand{\url}[1]{#1}
\csname url@samestyle\endcsname
\providecommand{\newblock}{\relax}
\providecommand{\bibinfo}[2]{#2}
\providecommand{\BIBentrySTDinterwordspacing}{\spaceskip=0pt\relax}
\providecommand{\BIBentryALTinterwordstretchfactor}{4}
\providecommand{\BIBentryALTinterwordspacing}{\spaceskip=\fontdimen2\font plus
\BIBentryALTinterwordstretchfactor\fontdimen3\font minus
  \fontdimen4\font\relax}
\providecommand{\BIBforeignlanguage}[2]{{%
\expandafter\ifx\csname l@#1\endcsname\relax
\typeout{** WARNING: IEEEtran.bst: No hyphenation pattern has been}%
\typeout{** loaded for the language `#1'. Using the pattern for}%
\typeout{** the default language instead.}%
\else
\language=\csname l@#1\endcsname
\fi
#2}}
\providecommand{\BIBdecl}{\relax}
\BIBdecl

\bibitem{3gpp-TR-36-885}
\emph{Study on LTE-based V2X services (v14.0.0, Release 14)}, 3GPP, July 2016,
  3GPP TR 36.885.

\bibitem{3gpp-rel16}
\emph{Release 16 Description; Summary of Rel-16 Work Items (V1.0.0 , Release
  16))}, 3GPP, Dec 2020, 3GPP TR 21.916.

\bibitem{limeric_paper}
G.~{Bansal}, J.~B. {Kenney}, and C.~E. {Rohrs}, ``Limeric: A linear adaptive
  message rate algorithm for dsrc congestion control,'' \emph{IEEE Transactions
  on Vehicular Technology}, vol.~62, no.~9, pp. 4182--4197, 2013.

\bibitem{Rostami2016}
A.~Rostami, S.~Member, B.~Cheng, S.~Member, G.~Bansal, K.~Sjoberg, M.~Gruteser,
  and J.~B. Kenney, ``{Channel Congestion Control Approaches},'' vol.~17,
  no.~10, pp. 1--14, 2016.

\bibitem{Kenney2013}
J.~B. Kenney and G.~Bansal, ``{Controlling Congestion in Safety-Message
  Transmissions},'' no. December, pp. 1--8, 2013.

\bibitem{cc-hari-missing-sci}
A.~{Mansouri}, V.~{Martinez}, and J.~{Härri}, ``A first investigation of
  congestion control for lte-v2x mode 4,'' in \emph{2019 15th Annual Conference
  on Wireless On-demand Network Systems and Services (WONS)}, 2019, pp. 56--63.

\bibitem{Wendland2019}
P.~Wendland, G.~Schaefer, and R.~S. Thoma, ``{LTE-V2X Mode 4: Increasing
  robustness and DCC compatibility with reservation splitting},'' \emph{2019
  8th IEEE International Conference on Connected Vehicles and Expo, ICCVE 2019
  - Proceedings}, pp. 0--5, 2019.

\bibitem{ETSI2018}
ETSI, ``{ETSI TS 102 687 - V1.2.1 - Intelligent Transport Systems ( ITS );
  Decentralized Congestion Control Mechanisms for Intelligent Transport Systems
  operating in the 5 GHz range ;},'' \emph{ETSI standard}, vol.~2, pp. 1--14,
  2018.

\bibitem{Amador2020}
O.~Amador, I.~Soto, M.~Uruena, and M.~Calderon, ``{GoT: Decreasing DCC queuing
  for CAM messages},'' \emph{IEEE Communications Letters}, vol.~24, no.~12, pp.
  2974--2978, 2020.

\bibitem{ETSI-cellular}
{ETSI TS 103 574 V0.3.1}, ``{Intelligent Transport Systems (ITS); Access layer
  part; Congestion Control for the Cellular: V2X PC5 interface},'' vol.~1, pp.
  1--12, 2018.

\bibitem{Qualcomm2016}
Qualcomm, ``{R1-1611594 Congestion control for V2V},'' Tech. Rep., 2016, 3GPP
  TSG-RAN WG1 No.87.

\bibitem{Mansouri2019}
A.~Mansouri, V.~Martinez, and J.~Harri, ``{A First Investigation of Congestion
  Control for LTE-V2X Mode 4},'' \emph{2019 15th Annual Conference on Wireless
  On-demand Network Systems and Services, WONS 2019 - Proceedings}, pp. 56--63,
  2019.

\bibitem{Sepulcre2020}
M.~Sepulcre, J.~Mira, G.~Thandavarayan, and J.~Gozalvez, ``{Is Packet Dropping
  a Suitable Congestion Control Mechanism for Vehicular Networks?}'' in
  \emph{IEEE Vehicular Technology Conference}, vol. 2020-May.\hskip 1em plus
  0.5em minus 0.4em\relax Institute of Electrical and Electronics Engineers
  Inc., may 2020.

\bibitem{BazziDoICare2018}
A.~Bazzi, G.~Cecchini, B.~M. Masini, and A.~Zanella, ``{Should i Really Care of
  That CAM?}'' \emph{IEEE International Symposium on Personal, Indoor and
  Mobile Radio Communications, PIMRC}, vol. 2018-Septe, 2018.

\bibitem{SAE-standard}
S.~o. A.~E. SAE~International, ``{On-board system requirements for v2v safety
  communications. Standard Doc J2945/1},'' 03 2016.

\bibitem{supra}
Y.~P. Fallah, N.~Nasiriani, and H.~Krishnan, ``Stable and fair power control in
  vehicle safety networks,'' \emph{IEEE Transactions on Vehicular Technology},
  vol.~65, no.~3, pp. 1662--1675, 2016.

\bibitem{toghi-dcc-first}
B.~Toghi, M.~Saifuddin, Y.~P. Fallah, and M.~O. Mughal, ``{Analysis of
  distributed congestion control in cellular vehicle-To-everything networks},''
  \emph{arXiv}, 2019.

\bibitem{Toghi-dcc-spatio}
B.~Toghi, M.~Saifuddin, M.~O. Mughal, and Y.~P. Fallah, ``{Spatio-temporal
  dynamics of cellular V2X communication in dense vehicular networks},''
  \emph{arXiv}, 2019.

\bibitem{YoonPowerRate}
Y.~Yoon and H.~Kim, ``{Balancing Power and Rate Control for Improved Congestion
  Control in Cellular V2X Communication Environments},'' \emph{IEEE Access},
  vol.~8, pp. 105\,071--105\,081, 2020.

\bibitem{ChoiQOS}
J.~Choi and H.~Kim, ``{A QoS-Aware Congestion Control Scheme for C-V2X Safety
  Communications},'' \emph{IEEE Vehicular Networking Conference, VNC}, vol.
  2020-Decem, pp. 23--26, 2020.

\bibitem{Khan2018}
I.~Khan and J.~H{\"{a}}rri, ``{Integration Challenges of Facilities-Layer DCC
  for Heterogeneous V2X Services},'' \emph{IEEE Intelligent Vehicles Symposium,
  Proceedings}, vol. 2018-June, no.~Iv, pp. 1131--1136, 2018.

\bibitem{Khan2020}
M.~I. Khan, M.~Sepulcre, and J.~Harri, ``{Cooperative Wireless Congestion
  Control for Multi-Service V2X Communication},'' \emph{IEEE Intelligent
  Vehicles Symposium, Proceedings}, no.~Iv, pp. 1357--1363, 2020.

\bibitem{Delooz2020}
Q.~Delooz, R.~Riebl, A.~Festag, and A.~Vinel, ``{Design and Performance of
  Congestion-Aware Collective Perception},'' \emph{IEEE Vehicular Networking
  Conference, VNC}, vol. 2020-Decem, 2020.

\bibitem{mccarthy2021opencv2x}
B.~McCarthy, A.~Burbano-Abril, V.~R. Licea, and A.~O'Driscoll, ``Opencv2x:
  Modelling of the v2x cellular sidelink and performance evaluation for
  aperiodic traffic,'' \emph{arXiv}, 2021.

\end{thebibliography}

\end{document}